\documentclass[letterpaper, 10 pt, conference]{ieeeconf}
\IEEEoverridecommandlockouts
\usepackage{cite}
\usepackage{amsmath,amssymb,amsfonts}
\usepackage[english]{babel}
\usepackage{algorithmic}
\usepackage{graphicx}
\usepackage{textcomp}
\usepackage{optidef}
\usepackage{listings}
\usepackage[ruled,vlined]{algorithm2e}
\newtheorem{theorem}{Theorem}
\newtheorem{corollary}{Corollary}
\newtheorem{assumption}{Assumption}

\newtheorem{defn}{Definition}
\newtheorem{ex}{Example}

\usepackage{caption}
\usepackage{subcaption}

\usepackage{placeins}
\usepackage{mathtools}
\usepackage{lipsum}
\usepackage{braket}

\usepackage{bbding}
\usepackage{pifont}
\usepackage{wasysym}
\usepackage{amssymb}
\usepackage{multirow}

\newcommand{\reviewer}[1]{}

\newcommand{\Rspace}{\mathbb{R}}

\newcommand{\id}{\textbf{I}}

\newcommand{\Ss}{\mathcal{S}}

\newcommand{\X}{\mathcal{X}}

\newcommand{\Z}{\mathcal{Z}}
\newcommand{\U}{\mathcal{U}}
\newcommand{\R}{\mathcal{R}}
\newcommand{\Pp}{\mathcal{P}}
\newcommand{\V}{\mathcal{V}}

\newcommand{\W}{\mathcal{W}}

\newcommand{\DPhi}{\text{D}_{\Phi}}

\newcommand{\suc}{\text{Suc}}

\newcommand{\MarginNotes}[1]{}

\usepackage{soul}

\def\BibTeX{{\rm B\kern-.05em{\sc i\kern-.025em b}\kern-.08em
    T\kern-.1667em\lower.7ex\hbox{E}\kern-.125emX}}
\begin{document}
\title{Set-valued State Estimation for Nonlinear Systems Using Hybrid Zonotopes}%
\author{Jacob A. Siefert, Andrew F. Thompson, Jonah J. Glunt, and Herschel C. Pangborn%
\thanks{Jacob A. Siefert, Andrew F. Thompson, Jonah J. Glunt, and Herschel C. Pangborn are with the Department of Mechanical Engineering, The Pennsylvania State University, University Park, PA 16802 USA (e-mail: jas7031@psu.edu; thompson@psu.edu; jglunt@psu.edu; hcpangborn@psu.edu).}}

\maketitle
\thispagestyle{empty} 

\begin{abstract}
This paper proposes a method for set-valued state estimation of nonlinear, discrete-time systems. This is achieved by combining \emph{graphs of functions} representing system dynamics and measurements with the \emph{hybrid zonotope} set representation that can efficiently represent nonconvex and disjoint sets. Tight over-approximations of complex nonlinear functions are efficiently produced by leveraging special ordered sets and neural networks, which enable computation of set-valued state estimates that grow linearly in memory complexity with time. A numerical example demonstrates significant reduction of conservatism in the set-valued state estimates using the proposed method as compared to an idealized convex approach.
\end{abstract}
\section{Introduction}
\label{sect:introduction}

Estimation methods are often needed to  implement state-feedback controllers~\cite{Simon2006} or detect faults and failures within a system~\cite{Raimondo2016faulDet}. Approaches for state estimation include stochastic methods such as Kalman filtering~\cite{Simon2006} and set-based methods~\cite{Rego2021,LOUKKAS20176471,WANG2018435}. Stochastic approaches have been successfully applied to a wide range of systems due to their efficient implementation and ability to provide a statistical characterization of the estimate uncertainty. This is contingent upon accurate representation of random processes, e.g., measurement noise. Statistical methods do not seek to provide guaranteed bounds on estimate uncertainties. 

This paper addresses set-valued state estimation (SVSE) methods, also referred to as set-membership state estimation, which compute sets containing \textit{all} possible state values consistent with a dynamic plant model, output measurements, and uncertainty bounds. SVSE methods commonly use convex set representations, such as zonotopes and constrained zonotopes~\cite{Althoff2021_linear,Le_Zono_SE2031, Combastel2005,scott2016constrained,Rego2018, Rego2020, Rego2021}, as these are closed under key set operations for propagating linear dynamics~\cite{Althoff2021_linear}.
However, convex sets are \textit{not} closed under nonlinear mappings. To address a larger class of dynamics and measurement models, a common approach is to generate linear approximations, e.g., via mean value and first-order Taylor extensions~\cite{Rego2020, Rego2021}.

Convex approaches are successful when the dynamics are represented well using affine approximations and when the exact set of states consistent with a given measurement can be tightly approximated with a convex set. When the true set of states consistent with the dynamic model, measurements, and uncertainty is significantly nonconvex, over-approximations using affine and convex approaches incur significant error. Methods considering collections of convex sets coupled with piecewise-affine approximations have been developed, though computational challenges arise, for example successive intersection with guards can result in exponential growth in the number of convex sets with time \cite{Althoff2021_linear}. It is especially challenging to compute successor sets used in the dynamic update of nonconvex SVSE when the set-valued state estimate is large \cite{Althoff2021_linear}. Similar difficulties arise in the measurement update.

\textit{Contribution:} This paper proposes a new approach for SVSE of nonlinear systems. Graphs of functions are represented with a nonconvex set representation called the hybrid zonotope, providing set estimates tighter to the true state than possible using convex methods while admitting efficient and scalable calculations. To the authors' knowledge, this is the first work to apply hybrid zonotopes for SVSE.

\textit{Outline:} 
Section~\ref{sec:Prelims} provides notation, reviews key set operations, and states fundamental definitions including that of the hybrid zonotope. Section~\ref{sect:SUS_reach} introduces SVSE and the key set identities of the proposed method. Section~\ref{sect:HZ_SVSE} describes how hybrid zonotopes can be used for nonconvex SVSE of nonlinear systems and compares methods for approximating unary and binary nonlinear functions. A numerical example in Section~\ref{sect:Examples} demonstrates the proposed method.
\section{Preliminaries and Previous Work}
\label{sec:Prelims}

\subsection{Notation}
Matrices are denoted by uppercase letters, e.g., $G\in\Rspace^{n\times n_g}$, and sets by uppercase calligraphic letters, e.g., $\mathcal{Z}\subset\Rspace^{n}$. Vectors and scalars are denoted by lowercase letters. The $i^{th}$ column of a matrix $G$ is denoted $G_{(\cdot,i)}$.
Commas in subscripts are used to distinguish between properties that are defined for multiple sets, e.g., $n_{g,z}$ and $n_{g,w}$ describe the complexity of the representation of $\mathcal{Z}$ and $\mathcal{W}$, respectively. The $n$-dimensional unit hypercube is denoted by $\mathcal{B}_{\infty}^n=\left\{x\in\Rspace^{n}~|~\|x\|_{\infty}\leq1\right\}$. The set of all $n$-dimensional binary vectors is denoted by $\{-1,1\}^{n}$ and the interval set between a lower bound $b_{l}$ and an upper bound $b_{u}$ is denoted by $[b_{l},b_{u}]$. Matrices of all $0$ and $1$ elements are denoted by $\mathbf{0}$ and $\mathbf{1}$, respectively, of appropriate dimension, and $\id$ denotes the identity matrix. The $i^{th}$ row of the identity matrix is given by $e_i$. The concatenation of two column vectors to a single column vector is denoted by $(g_1,\:g_2)=[g_1^T\:g_2^T]^T$.

\subsection{Set Operations, Representations, \& Graphs of Functions}
\label{sec:HybZono}

Given the sets $\mathcal{Z},\mathcal{W}\subset\Rspace^{n},\:\mathcal{Y}\subset\Rspace^{m}$, and matrix $R\in\Rspace^{m\times n}$, the linear transformation of $\mathcal{Z}$ by $R$ is $R\mathcal{Z}=\{Rz~|~z\in\mathcal{Z}\}$, the Minkowski sum of $\mathcal{Z}$ and $\mathcal{W}$ is $\mathcal{Z}\oplus\mathcal{W}=\{z+w~|~z\in\mathcal{Z},\:w\in\mathcal{W}\}$, the generalized intersection of $\mathcal{Z}$ and $\mathcal{Y}$ under $R$ is $\mathcal{Z}\cap_R\mathcal{Y}=\{z\in\mathcal{Z}~|~Rz\in\mathcal{Y}\}$, and the Cartesian product of $\mathcal{Z}$ and $\mathcal{Y}$ is $\mathcal{Z}\times\mathcal{Y}=\{(z,y)|~z\in\mathcal{Z},\:y\in\mathcal{Y}\}$. 

\begin{defn}
    \label{def-VrepPoly}
    A set $\Pp\in\Rspace^n$ is a convex polytope if it is bounded and there exists $V\in\Rspace^{n \times n_v}$ with $n_v<\infty$ such that
    \begin{align}
    \nonumber
        \Pp = \left\{ V\lambda\ |\ \lambda_j\geq0\ \forall j\in \{1,...,n_v\},\ \mathbf{1}^T \lambda = 1\right\}\:.
    \end{align}
\end{defn}%

A convex polytope is the convex hull of a finite set of vertices given by the columns of $V$. A convex polytope defined using a matrix of concatenated vertices is said to be given in \emph{vertex representation} (V-rep). Constrained zonotopes are an alternative representation for polytopes with computational advantages for key set operations \cite{scott2016constrained}. Hybrid zonotopes extend constrained zonotopes by the addition of binary factors, allowing for representation of nonconvex and potentially disjoint sets while maintaining computational advantages for set operations associated with constrained zonotopes. The proposed method leverages the ability to efficiently represent nonconvexities within a single set, as opposed to via a collection of convex sets.

\begin{defn}\label{def-hybridZono} \cite[Def. 3]{Bird_HybZono}
The set $\mathcal{Z}_h\subset\Rspace^n$ is a \emph{hybrid zonotope} if there exists $G^c\in\Rspace^{n\times n_{g}}$, $G^b\in\Rspace^{n\times n_{b}}$, $c\in\Rspace^{n}$, $A^c\in\Rspace^{n_{c}\times n_{g}}$, $A^b\in\Rspace^{n_{c}\times n_{b}}$, and $b\in\Rspace^{n_c}$ such that {\small
    \begin{equation}\label{def-eqn-hybridZono}
        \mathcal{Z}_h = \left\{ \left[G^c \: G^b\right]\left[\begin{smallmatrix}\xi^c \\ \xi^b \end{smallmatrix}\right]  + c\: \middle| \begin{matrix} \left[\begin{smallmatrix}\xi^c \\ \xi^b \end{smallmatrix}\right]\in \mathcal{B}_\infty^{n_{g}} \times \{-1,1\}^{n_{b}}, \\ \left[A^c \: A^b\right]\left[\begin{smallmatrix}\xi^c \\ \xi^b \end{smallmatrix}\right] = b \end{matrix} \right\}\:.
\end{equation}}
\vskip \baselineskip
\end{defn}

 A hybrid zonotope is the union of $2^{n_b}$ constrained zonotopes corresponding to the possible combinations of binary factors $\xi^b$. The hybrid zonotope \eqref{def-eqn-hybridZono} is given in shorthand notation as $\mathcal{Z}_h=\langle G^c,G^b,c,A^c,A^b,b\rangle\subset\Rspace^n$. Identities and time complexity scaling of linear transformations, Minkowski sums, generalized intersections, and generalized half-space intersections are reported in \cite[Section 3.2]{Bird_HybZono}. An identity and time complexity for Cartesian products are given in \cite{BIRDthesis_2022}. Methods for removing redundant generators and constraints of a hybrid zonotope were reported in \cite{Bird_HybZono} and further developed in \cite{BIRDthesis_2022}.

A set-valued mapping $\phi: \DPhi \rightarrow\mathcal{Q}$ assigns each element $p\in\text{D}_{\Phi}$ to a subset of $\mathcal{Q}$. We refer to the set $\Phi = \{(p,q)\ |\ p\in\DPhi,\ q\in\phi(p)\subseteq\mathcal{Q} \}$ as a \emph{graph of the function} $\phi$. The set $\text{D}_{\Phi}$ is referred to as the \emph{domain set} of $\Phi$ and can be chosen by a user as the set of inputs of interest. The proposed methods in this paper over-approximate graphs of nonlinear functions with hybrid zonotopes by leveraging Special Ordered Set (SOS) approximations. 

\subsection{Special Ordered Sets}
\label{sec:SOSintro}

SOS approximations were originally developed to approximate solutions of nonlinear optimization programs by replacing nonlinear functions with piecewise-affine approximations \cite{beale1970_SOS}. In this section, we define the SOS approximation, which can be conveniently represented as a collection of V-rep polytopes, and provide an identity to convert a collection of V-rep polytopes into a hybrid zonotope.

\begin{defn}
An SOS approximation $\Ss$ of a scalar-valued function $f(x):\Rspace^n\rightarrow\Rspace$ is defined by a vertex matrix $V=[v_1,v_2,...,v_{n_v}]\in\Rspace^{(n+1)\times n_v}$ such that $v_i = (x_i,f(x_i))$ and is given by $\Ss=\{ V \lambda\ |\ \mathbf{1}^T \lambda = 1,\ \mathbf{0}\leq\lambda,$ where at most $n+1$ entries of $\lambda$ are nonzero and correspond to an $n$-dimensional simplex$\}.$
\end{defn}%
We refer to \emph{breakpoints} as the choice of $x_i, i\in \{1,...,n_v\}$ used to construct an SOS approximation.

\begin{theorem}
\label{thm-SOS2HYBZONO} \cite[Theorem 4]{Siefert2023_TAC}
Consider a set $\Ss$ consisting of the union of $N$ V-rep polytopes, $\Ss=\cup_{i=1}^{N}\Pp_i$, with a total of $n_v$ vertices. Define the vertex matrix $V=[v_1,\dots,v_{n_v}]\in\Rspace^{n\times n_v}$ and construct a corresponding incidence matrix $M\in\Rspace^{n_v \times N}$, with entries $M_{(j,i)}\in\{0,1\} \:\forall \:i,j$, such that 
    \begin{align}
    \nonumber
        \Pp_i=\left\{ V\lambda\ \bigg|\ \lambda_j \in \begin{cases} [0,1], & \forall\ j\in \{ k\ |\ M_{(k,i)}=1 \}\: \\ \{0\}, & \forall\ j\in \{ k\ |\ M_{(k,i)}=0 \} \end{cases} \right\}\:.
    \end{align}%
Define the hybrid zonotope
\begin{equation}\nonumber
    \mathcal{Q}=\frac{1}{2}\left\langle\begin{bmatrix}
        \mathbf{I}_{n_v} \\ \mathbf{0}
    \end{bmatrix},\begin{bmatrix}
        \mathbf{0} \\ \mathbf{I}_{N}
    \end{bmatrix},\begin{bmatrix}
        \mathbf{1}_{n_v} \\ \mathbf{1}_{N}
    \end{bmatrix}, \begin{bmatrix}
        \mathbf{1}_{n_v}^T\\ \mathbf{0}
    \end{bmatrix}, \begin{bmatrix}
        \mathbf{0}\\ \mathbf{1}_{N}^T
    \end{bmatrix},\begin{bmatrix}
        2-n_v \\ 2 -N
    \end{bmatrix}\right\rangle\:,
\end{equation}
the polyhedron $\mathcal{H}=\{h\in\Rspace^{n_v}~\vert~ h\leq\mathbf{0}\}$, and let
\begin{equation}\label{prop-SOS-eqn-basis}
    \mathcal{D}=\mathcal{Q}\cap_{[\mathbf{I}_{n_v}~-M]}\mathcal{H}\:.
\end{equation}
Then the set $\Ss$ is equivalently given by the hybrid zonotope
\begin{equation}\label{prop-SOS-eqn}
    \Z_{S}=\begin{bmatrix}
        V &\mathbf{0}
    \end{bmatrix}\mathcal{D}\:,
\end{equation}
with memory complexity $n_{g}=2 n_v,\ n_{b}=N,\ n_{c}=n_{v}+2$.
\end{theorem}

\subsection{Set-valued State Estimation}

Consider a class of discrete-time nonlinear systems given by dynamic plant model $f:\Rspace^{n_x}\times \Rspace^{n_u}\times\Rspace^{n_w}\rightarrow\Rspace^{n_x}$ and measurement model $g: \Rspace^{n_x}\times\Rspace^{n_v}\rightarrow\Rspace^{n_y}$,
\begin{align} 
\label{eqn-genNLdyn}
    x_{k+1} &= f(x_{k},u_{k},w_k)\:, \\
\label{eqn-genNLmeas}
    y_{k+1}& = g(x_{k+1},v_{k+1})\:,
\end{align}%
with bounded state $x\in\mathcal{X}\subset\Rspace^{n_x}$, input $u\in\mathcal{U}\subset\Rspace^{n_u}$, process uncertainty $w\in\mathcal{W}\subset\Rspace^{n_w}$, and measurement noise $v\in\mathcal{V}\subset\Rspace^{n_v}$.
\begin{assumption}
\label{as-bounded_fandg}
The functions $f(\cdot)$ and $g(\cdot)$ are defined and bounded over their respective domains corresponding to $\X$, $\U$, $\W$, and $\V$, which themselves are also bounded.
\end{assumption}%
Assumption \ref{as-bounded_fandg} is important for compatibility with hybrid zonotopes because they are inherently bounded.

The set $\hat{\mathcal{X}}_{k+1|k}$ denotes the set of states that can be achieved by the dynamics at time step $k+1$ that are consistent with all measurements through time step $k$ and may be found as the \emph{successor set} of $\hat{\mathcal{X}} _{k|k}$, $\suc(\hat{\mathcal{X}}_{k|k},\mathcal{U}_{k},\mathcal{W}_{k})$.

\begin{defn}
The \emph{successor set} from $\mathcal{R}_k\subseteq\mathcal{X}$ with inputs and process uncertainty bounded by $\U_k\subseteq\mathcal{U}$ and $\W_k\subseteq\mathcal{W}$, respectively, is given by
{\small
\begin{align}\label{eqn-Suc}
    \suc(\R_{k},\U_k,\W_k)\equiv\left\{\begin{matrix}
        f(x,u,w)
        \mid\:
        \begin{array}{c}
             x\in\mathcal{R}_{k},\: u\in\U_k\:,  \\
             w\in\W_k 
        \end{array}
    \end{matrix}
    \right\}\:.
\end{align}}%
\vskip \baselineskip
\end{defn}

\begin{defn}
    The set $\hat{\X}_{y_{k+1}}$ is  the set of states consistent with measurement $y_{k+1}$ and measurement noise $\mathcal{V}_{k+1} \subseteq \mathcal{V}$ within the feasible region $\mathcal{X}$ and is given by
    \begin{align}
        \label{eqn-MeasConsistentStates}
        \hat{\X}_{y_{k+1}} \equiv \left\{x\ |\ y_{k+1} = g(x,v),\ x\in\mathcal{X},\ v\in\mathcal{V}_{k+1} \right\} \:.
    \end{align}
\end{defn}

Assuming the initialized set-valued state estimate contains the true initial state, i.e., $x_0\in\hat{\X}_{0|0}$, SVSE combines dynamic and measurement updates using the recursion
\begin{align}
    \label{eqn-dynUpdate}
    \hat{\mathcal{X}}_{k+1|k} &= \suc(\hat{\mathcal{X}}_{k|k},\U_k,\W_k) \:,\\
    \label{eqn-measUpdate}
    \hat{\mathcal{X}}_{k+1|k+1} &=\hat{\mathcal{X}}_{k+1|k} \cap \hat{\mathcal{X}}_{y_{k+1}} \:,
\end{align}%
to generate set-valued state estimates guaranteed to contain the true state.
Section \ref{sect:SUS_reach} provides identities to calculate \eqref{eqn-dynUpdate}-\eqref{eqn-measUpdate} and Section \ref{sect:HZ_SVSE} efficiently implements the identities for nonlinear systems using hybrid zonotopes.
\section{Set Identities for Graphs of Functions}
\label{sect:SUS_reach}

Calculating the successor set of nonlinear systems using set propagation techniques has been addressed by a wide variety of methods and set representations~(see \cite{Althoff2021_linear} and references therein). Recent work by the authors has provided identities for calculating exact successor sets of certain classes of discrete-time hybrid systems (mixed logical dynamical systems and linear systems in closed loop with model predictive control) and over-approximations of successor sets for discrete-time nonlinear systems \cite{Siefert2023,SiefertHybSUS,  Siefert2023_TAC}. Computational complexity scales linearly with the dimension of the system and memory complexity grows linearly with time. This is achieved using a combination of the hybrid zonotope set representation and a construct called the state-update set, which is equivalent to the graph of the function describing a discrete-time system's dynamics.
We now generalize the state-update set identities first developed for forward and backward reachability to more general graphs of functions, allowing for computation of an ``output'' set given an ``input'' set and vice versa.

\begin{theorem}
\label{thm-in2out}
If the set of inputs satisfies $\mathcal{P}\subseteq\DPhi$, then the set of outputs, defined as
\begin{align}
    \label{eqn-def-Q}
    \mathcal{Q} \equiv \left\{ q\ |\ q\in\phi(p),\ p\in\mathcal{P} \right\}\:,
\end{align} can be calculated using the graph of the function $\phi$, $\Phi=\{(x,\phi(x))|x\in\text{D}_\Phi\}$, as
\begin{align}
    \label{eqn-ID-Q}
    \mathcal{Q} = \begin{bmatrix} \mathbf{0} & \id_{n_q} \end{bmatrix} \left( \Phi \cap_{\begin{bmatrix}
        \id_{n_p} & \mathbf{0}
    \end{bmatrix}} \mathcal{P} \right)\:.
\end{align}
\begin{proof}
By definitions of generalized intersections and linear transformations
{\small
\begin{align}
\nonumber
    \begin{bmatrix} \mathbf{0} & \id_{n_q} \end{bmatrix} \left( \Phi \cap_{\begin{bmatrix}
        \id_{n_p} & \mathbf{0}
    \end{bmatrix}} \mathcal{P} \right) = \left\{ q\ |\ q\in\phi(p),\ p\in\mathcal{P}\cap\DPhi\right\}\:.
\end{align}}%
The assumption $\mathcal{P}\subseteq\DPhi \implies \mathcal{P}\cap\DPhi = \mathcal{P}$.
\end{proof}
\end{theorem}

\begin{theorem}
\label{thm-out2in}
The set of inputs $\mathcal{P}_{\DPhi}$ within a region of interest $\DPhi$ and consistent with the output set $\mathcal{Q}$, defined by
\begin{align}
    \label{eqn-def-P}
    \mathcal{P}_{\DPhi} \equiv \left\{ p\ |\ q\in\phi(p),\ q\in\mathcal{Q},\ p\in\DPhi  \right\}\:,
\end{align} can be calculated using the graph of the function $\phi$, $\Phi=\{(x,\phi(x))|x\in\text{D}_\Phi\}$, as
\begin{align}
    \label{eqn-ID-P}
    \mathcal{P}_{\DPhi} = \begin{bmatrix} \id_{n_p} & \mathbf{0} \end{bmatrix} \left( \Phi \cap_{\begin{bmatrix}
         \mathbf{0} & \id_{n_q}
    \end{bmatrix}} \mathcal{Q} \right)\:.
\end{align}
\begin{proof}
By applying the definitions of generalized intersections and linear transformations, the right side of \eqref{eqn-ID-P} yields \eqref{eqn-def-P}.
\end{proof}
\end{theorem}

Table \ref{tab:IOStypes} demonstrates how \textbf{Theorem \ref{thm-in2out}} and \textbf{Theorem \ref{thm-out2in}} are used to calculate the successor sets used in the dynamic update and the measurement-consistent state set $\hat{\mathcal{X}}_{y_{k+1}}$. Multiple argument vectors may be concatenated to match the single argument used in \textbf{Theorem \ref{thm-in2out}} and \textbf{Theorem \ref{thm-out2in}}, e.g., for the successor set $p=(x_k,u_k,w_k)$.

\begin{table}[htb!]
    \centering
    \caption{Identities for state estimation.}
    \label{tab:IOStypes}
    \begin{tabular}{l||l|l|c}
         & $q = \phi(p)$ & $\DPhi$ & Theorem  \\ \hline
         \hline
         $\suc(\hat{\mathcal{X}}_{k|k})$ & $x_{k+1}=f(x_k,u_k,w_k)$ & $\mathcal{X}\times \mathcal{U}\times \mathcal{W}$ &\ref{thm-in2out} \\
         $\hat{\mathcal{X}}_{y_{k+1}}$ & $y_{k+1}=g(x_{k+1},v_{k+1})$ & $\mathcal{X}\times\mathcal{V}$ & \ref{thm-out2in} 
    \end{tabular}
\end{table}

A fundamental challenge of reachability analysis is that a given set representation and its operations can only achieve efficient computation of \emph{exact} successor sets for a limited class of system models \cite{Gan_RA4SolvableDynSys}. SVSE inherits this challenge and the similar challenge of exactly representing the set of states consistent with a measurement model $\hat{\mathcal{X}}_{y_{k+1}}$. To obtain formal guarantees for a broader class of systems and measurement models \eqref{eqn-genNLdyn}-\eqref{eqn-genNLmeas}, \emph{over-approximations} of $\suc(\hat{\mathcal{X}}_{k|k})$ and $\hat{\mathcal{X}}_{y_{k+1}}$ can be computed instead. To this end, \textbf{Corollary~\ref{co-IOS-IDs4OA}} extends the previous results by considering the effect of using an over-approximation of the graph of a function.

\begin{corollary} \label{co-IOS-IDs4OA} For the identities \eqref{eqn-ID-Q} and \eqref{eqn-ID-P} provided by \textbf{Theorem \ref{thm-in2out}} and \textbf{Theorem \ref{thm-out2in}}, respectively, if $\Phi$ is replaced with an over-approximation $\Bar{\Phi}$, then the identities will instead yield over-approximations of the left sides.

\begin{proof}
    Set containment is preserved under linear transformation and generalized intersection.
\end{proof} 
\end{corollary}

Using over-approximations of $\suc(\hat{\mathcal{X}}_{k|k},\U_k,\W_k)$ and/or $\hat{\mathcal{X}}_{y_{k+1}}$ retains the guarantee that the recursion in \eqref{eqn-dynUpdate}-\eqref{eqn-measUpdate} calculates sets containing the true state, although it is desirable to minimize error in the over-approximations to achieve tight uncertainty bounds on the state estimate.
\section{Graphs of Functions via Hybrid Zonotopes}
\label{sect:HZ_SVSE}
\subsection{Motivation for Using Hybrid Zonotopes}
The identities in \eqref{eqn-ID-Q} and \eqref{eqn-ID-P} use linear transformations and generalized intersections, therefore the chosen set representation should be closed and ideally computationally efficient and scalable for these operations. Furthermore, the set representation must be able to represent the graph of dynamic and measurement functions. When \eqref{eqn-genNLdyn} and/or \eqref{eqn-genNLmeas} are nonlinear, it becomes important that the set representation can capture nonconvexities and discontinuities inherent to their graphs. In addition to hybrid zonotopes, other set representations such as constrained polynomial zonotopes \cite{kochdumper2023constrained} and sublevel sets exhibit these properties. The interested reader is referred to \cite{Althoff2021_linear} and references therein for alternative set representations. 

The remainder of this paper uses hybrid zonotopes for three reasons. Firstly, this results in linear growth in memory complexity, whereas representing nonconvex sets using a collection of convex sets would result in exponential growth in memory complexity \cite{Althoff2021_linear}. Specifically, the identities in \eqref{eqn-ID-Q} and \eqref{eqn-ID-P} of \textbf{Theorems \ref{thm-in2out}} and \textbf{\ref{thm-out2in}}, respectively, result in memory complexity 
{\begin{align}
\footnotesize
        \nonumber
        \begin{split}
    &\text{\underline{\textbf{Theorem~\ref{thm-in2out}}}}\\ 
        n_{g,{q}} &= n_{g,p}+n_{g,\phi}\:,\\
         \nonumber
        n_{b,{q}} &= n_{b,p}+n_{b,\phi}\:,\\
         \nonumber
        n_{c,{q}} &= n_{c,p}+n_{c,\phi}+n\:,
        \end{split}
        \footnotesize
        \nonumber
        \begin{split}
        &\text{\underline{\textbf{Theorem~\ref{thm-out2in}}}}\\
        n_{g,{p}} &= n_{g,q}+n_{g,\phi}\:,\\
         \nonumber
        n_{b,{p}} &= n_{b,q}+n_{b,\phi}\:,\\
         \nonumber
        n_{c,{p}} &= n_{c,q}+n_{c,\phi}+n\:.
        \end{split}
\end{align}}%
The intersection in \eqref{eqn-measUpdate} also results in linear memory complexity growth, therefore the set-valued state estimates exhibit linear memory growth with each time step when hybrid zonotopes are used. Secondly, the computational complexities of \eqref{eqn-ID-Q} and \eqref{eqn-ID-P} scale linearly with $n_p$ and $n_q$, which relate to the dimensions of the state, control input, process uncertainty, measurement, and measurement noise. Thirdly, existing methods for piecewise-affine approximations of nonlinear functions using hybrid zonotopes can be leveraged. Such methods include SOS approximations \cite{Siefert2023}, unions of polytope enclosures \cite{Siefert2023_TAC}, and neural networks with rectified linear unit activation functions \cite{Siefert2023_TAC}.

The efficient representation of nonconvex sets enables tight over-approximation of graphs of nonlinear functions. Identities to generate over-approximations of unary and binary nonlinear functions as hybrid zonotopes and then compose them to capture multivariate nonlinear functions are presented in \cite{Siefert2023_TAC}. This shows how to construct over-approximations of $\Phi$ that avoid the so-called \emph{curse of dimensionality}, with polynomial rather than exponential memory complexity as a function of the state dimension. However, this prior work only considered approximations of unary and binary functions with uniformly spaced breakpoints.

\subsection{Approximations of Unary and Binary Functions}

This section presents and compares three methods for over-approximating graphs of nonlinear functions over a domain using hybrid zonotopes. The first two methods consider the placement of a fixed number of \emph{breakpoints} $x_{break,i}, \forall i\in\{1,...,n_{v}\}$ used to generate SOS approximations, which can then be represented as a collection of V-rep polytopes and converted to a hybrid zonotope using \textbf{Theorem \ref{thm-SOS2HYBZONO}}. Method M1 uses a uniform spacing of breakpoints and serves as a benchmark for the other methods. Method M2 attempts to place the locations of breakpoints optimally to minimize the maximum error between the SOS approximation $\psi(\cdot)$ and the nonlinear function $\phi(\cdot)$,
\begin{align}
\label{eqn-XbreakOpt}
    \min_{x_{break}} \max_{x}\ & |\psi(x)-\phi(x)|\\
    \nonumber
    \text{s.t.}\ &x\in\text{D}_{\Phi} \:.
\end{align}%
The optimization program \eqref{eqn-XbreakOpt} is solved using MATLAB's ``fmincon" function.

Method M3 trains a fully connected neural network using the rectified linear unit (ReLU) activation function, implemented with the MATLAB Deep Learning Toolbox. The neural network is then represented exactly as a hybrid zonotope using the methods from \cite{Siefert2023_TAC}. 

All three methods produce over-approximations by taking a Minkowski sum with a set that bounds the maximum approximation error. Example \ref{ex-H-inv} and Example \ref{ex-H-Towers} compare these methods.

\begin{ex}
\label{ex-H-inv}
Consider the function $y=\frac{1}{x}$ on the domain $x\in[1,10]$. M1 and M2 use 5 breakpoints and M3 uses a single hidden layer with 4 nodes that results in similar memory complexity to M1 and M2. Table \ref{tab-H-inv-complexity} states the memory complexity of each method, and Figure \ref{fig-H-inv} plots each over-approximation. For the neural network of M3, only 2 nodes activate both regions of the ReLU function (M3 has only 2 binary generators) indicating unused ReLU activation functions. Comparing M1 to both M2 and M3 shows that a nonuniform spacing of piecewise-affine approximations can reduce error of the SOS approximation.

\begin{table}[htb!]
\caption{Memory complexity for set-based over-approximations of $y=\frac{1}{x}$.}
    \centering
    \begin{tabular}{l|c|c|c}
        Method & $n_{g}$ & $n_{b}$ & $n_{c}$  \\ \hline
        M1 (5 breakpoints) & 12 & 5 & 7\\\hline
        M2 (5 breakpoints) & 12 & 5 & 7\\\hline
        M3 (1 layer, 4 nodes) & 14 & 2 & 8\\
    \end{tabular}
    \label{tab-H-inv-complexity}
    \vspace*{-\baselineskip}
\end{table}

\begin{figure}[htb!]
    \centering
    \includegraphics[width=3in]{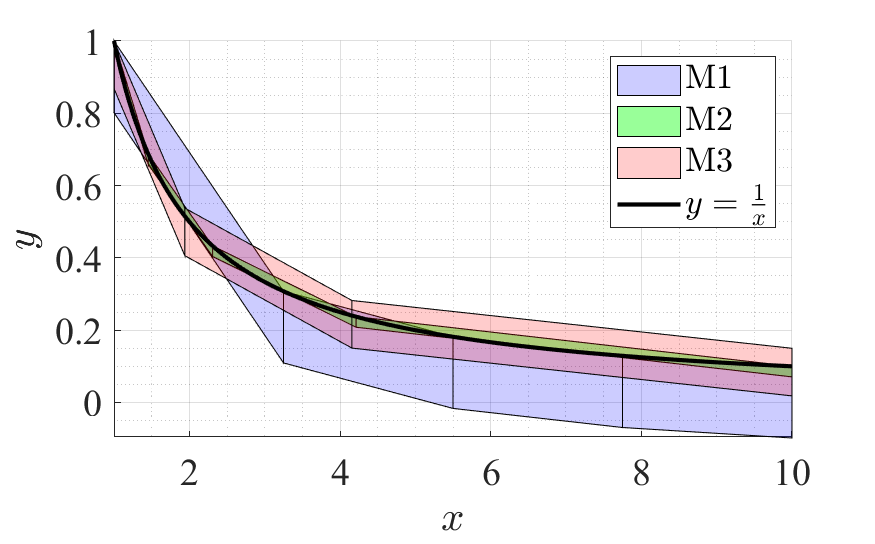}
    \caption{Comparison of methods for hybrid zonotope over-approximations of $y=\frac{1}{x}$. M1: uniformly spaced breakpoints, M2: optimized breakpoints per \eqref{eqn-XbreakOpt}, and M3: trained neural network.}
    \label{fig-H-inv}
\end{figure}
\end{ex}

\begin{ex}
\label{ex-H-Towers}
Consider the function
\begin{align}
\label{eqn-y-source}
    y = \sum_{i=1}^4 \frac{1}{d_i^2+1}\:,
\end{align}
where $d_i=||x-s_i||_2$ and {\small
\begin{equation}
    s_1 = (1,3),\ s_2 = (-2,2),\ s_3 = (3,0),\ s_4 = (-1,-4)\:. 
\end{equation}}%
One interpretation is that $y$ is a measurement of the summation of signal strengths emitted by four sources located at $s_i, \forall i\in\{1,2,3,4\}$, where the strength of signal from each source decreases with the inverse square of the distance from the source. Figure \ref{fig-H-tower} shows M1 and M3 over-approximations of \eqref{eqn-y-source} over a domain of $(x_1,x_2)\in[-5, 5]\times[-5, 5]$. M1 uses 100 breakpoints, while M3 uses 2 layers with 20 nodes each. M2 is omitted in this example, as formulating and solving an optimization program to select not only the location of breakpoints, but also the desired tessellation, falls outside the scope of this paper. Table \ref{tab-H-sources-complexity} and Figure \ref{fig-H-tower} show that M3 can achieve a much tighter approximation with less than a quarter of the maximum error of M1 and considerably lower memory complexity.

\begin{table}[htb!]
\caption{Memory complexity for set-based over-approximations of \eqref{eqn-y-source}.}
    \centering
    \begin{tabular}{l|c|c|c}
        Method & $n_{g}$ & $n_{b}$ & $n_{c}$  \\ \hline
        M1 (100 breakpoints) & 202 & 163 & 102\\\hline
        M3 (2 layers, 20 nodes each) & 127 & 31 & 93\\
    \end{tabular}
    \label{tab-H-sources-complexity}
    \vspace*{-\baselineskip}
\end{table}

\begin{figure}[htb!]
    \centering
     \begin{subfigure}[b]{3in}
         \centering
         \includegraphics[width=2.65 in, trim=1cm 1cm 0 0]{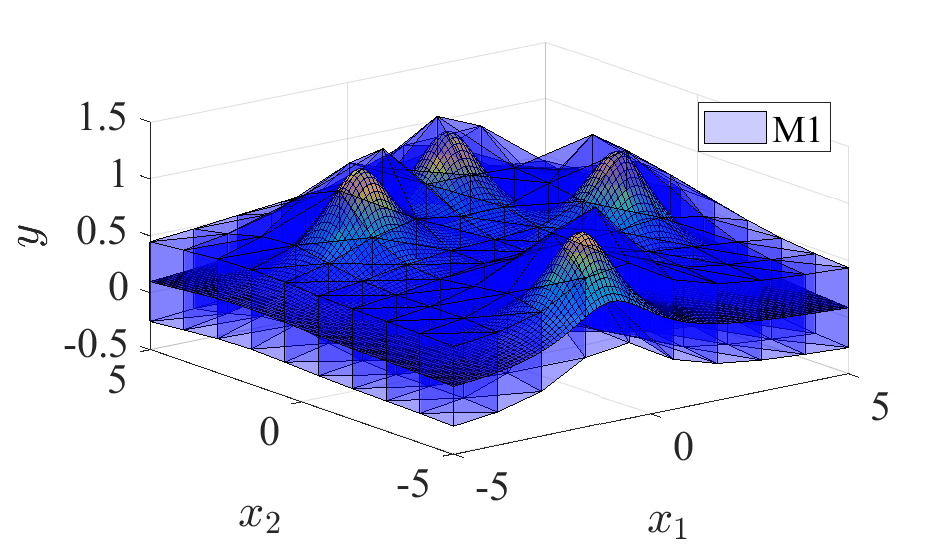}
         \caption{}
         \label{fig:TowerH_M1}
     \end{subfigure}
      \begin{subfigure}[b]{3in}
         \centering
         \includegraphics[width=2.65 in, trim=1cm 1cm 0 0]{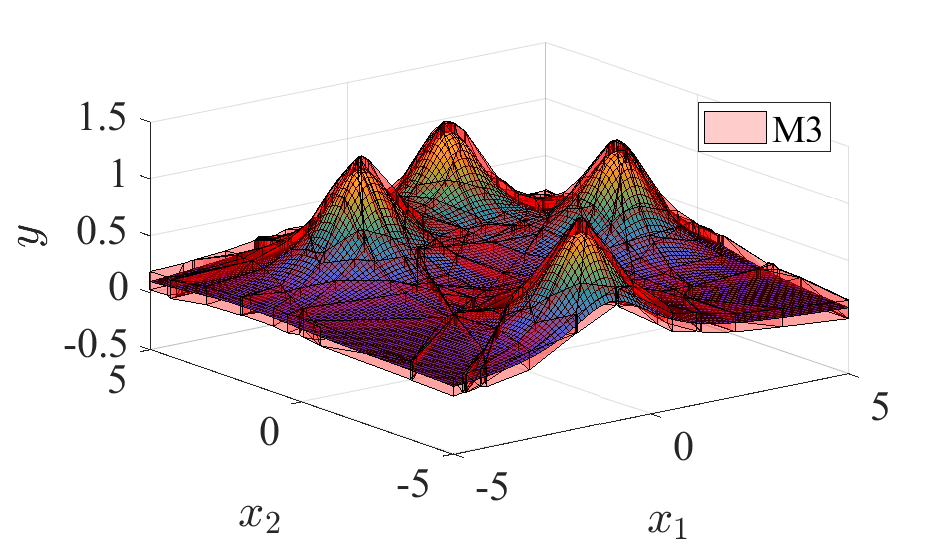}
         \caption{}
         \label{fig:TowerH_M3}
     \end{subfigure}
    \caption{Comparison of methods for hybrid zonotope over-approximation of \eqref{eqn-y-source}. M1: uniformly spaced breakpoints and M3: trained neural network. A surface plot of \eqref{eqn-y-source} is also shown in both (a) and (b).}
    \label{fig-H-tower}
\end{figure}
\end{ex}
\section{Numerical Example}
\label{sect:Examples}
Results in this section were generated with MATLAB on a desktop computer with a 3.0 GHz Intel i7 processor and 16 GB of RAM. Set-valued state estimates were plotted using techniques from \cite{Bird_HybZono} and \cite{BIRDthesis_2022}. 

Consider the integrator dynamics
\begin{align}
    x_{k+1} = x_k + u_k\:,
\label{eq:integrator}\end{align}
with $x_k,u_k\in\Rspace^2$, together with the nonlinear measurement equation given by \eqref{eqn-y-source}. While the proposed methods account for input uncertainty, process uncertainty, and measurement noise, for simplicity of exposition this example omits these. This allows us to more clearly assess the effect of using an over-approximated graph of a nonlinear function (see \textbf{Corollary \ref{co-IOS-IDs4OA}}). Method M3 from Example \ref{ex-H-Towers}, shown in Figure \ref{fig-H-tower}(b), serves as an over-approximation of a graph of a function $\Bar{\Phi}_{meas}$ for the measurement model. 

For this example, the dynamic update \eqref{eqn-dynUpdate} specifies to
\begin{align}
\label{eqn-numex-dynUpdate}
\hat{\mathcal{X}}_{k+1|k} &= \hat{\mathcal{X}}_{k|k} \oplus \{u_k\} \:,
\end{align}%
where $u_k$ is known to the estimator. The set $\hat{\mathcal{X}}_{y_{k+1}}$ in \eqref{eqn-MeasConsistentStates} is calculated using \textbf{Theorem \ref{thm-out2in}}, \textbf{Corollary \ref{co-IOS-IDs4OA}}, and $\Bar{\Phi}_{meas}$ as
\begin{align}
\nonumber
    \hat{\mathcal{X}}_{y_{k+1}} = \begin{bmatrix} \id_{2} & \mathbf{0} \end{bmatrix} \left( \Bar{\Phi}_{meas} \cap_{\begin{bmatrix}
         \mathbf{0} & 1
    \end{bmatrix}} \{y_{k+1}\} \right)\:,
\end{align}%
for use in \eqref{eqn-measUpdate}.
The set-valued estimate is initialized as the interval set corresponding to state constraints \begin{align}
    \label{eqn-numex-initialization}
    \hat{\mathcal{X}}_{0|-1} = \X =  [-5,5]\times[-5,5]\:,
\end{align}%
and the true initial state and input trajectory are given by
\begin{align}
    x_0 &= \begin{bmatrix} 1 & 0 \end{bmatrix}^T\:,\\
    [u_0\ u_1\ u_2\ u_3] &= \begin{bmatrix}
    -1 & -2 & -1 & 2\\
    1 & -1 & -1 & -1
    \end{bmatrix}\:.
\end{align}

Figure \ref{fig:SVSEofSources} shows the set-valued \textit{a~priori} state estimate $\Bar{\hat{\mathcal{X}}}_{k|k-1}$, \textit{a~posteriori} state estimate $\Bar{\hat{\mathcal{X}}}_{k|k}$,  exact set of measurement-consistent states $\hat{\mathcal{X}}_{y_{k}}$ and its over-approximation $\Bar{\hat{\mathcal{X}}}_{y_{k}}$, and true state $x_k$ for several time steps. This shows how the initial state estimate is gradually reduced to a small region by combining knowledge of the dynamics and measurement, and highlights how the nonlinearity of the measurement model is tightly captured by the nonconvexity of the hybrid zonotope set representation. When $k=2$, the state estimate consists of several disjoint regions, and after the third measurement, the state estimate consists of only two disjoint regions. After the fourth dynamic update, one of these regions escapes the set of feasible states, so only one region remains.

For comparison, a convex approach M$_{convex}$ is shown that estimates $\hat{\mathcal{X}}_{k|k}$ using convex methods by taking an inner convex approximation of $\hat{\mathcal{X}}_{y_{k}}$.
This is achieved using the same dynamic update \eqref{eqn-numex-dynUpdate} and initialization \eqref{eqn-numex-initialization}, but samples the exact level set of \eqref{eqn-y-source} within $\X$ at each time step to generate a polytope in vertex representation to approximate $\hat{\X}_{y_{k+1}}$.
It can be seen from M$_{convex}$ that \emph{any} convex approach would produce a much larger over-approximation than the proposed approach, which leverages nonconvex representations of $\hat{\X}_{y_{k+1}}$.

Set-valued state estimates for $k\in\{0,1,...,4\}$ are computed by the proposed method in less than 0.2 seconds. Plotting required 740 seconds as this involves solving mixed-integer linear programs to find each nonempty convex set within the hybrid zonotope, associated with specific values of the binary factors. However, upper and lower bounds on each of the states can be calculated much more quickly by sampling the support function in the axis-aligned directions. For this example, these bounds were obtained  for all time steps in a total of 6 seconds. 

In some applications it may be desirable to obtain a single point within the set-valued set estimate, for example as state feedback for a controller. 

This becomes nontrivial for nonconvex and/or disjoint set-valued state estimates, for which the ``center'' of a set may not be within the set, such as a toroid. A natural alternative would be to find a single feasible point within the set, which in this case can be found by solving a mixed-integer linear feasibility problem. This required $0.8$ seconds for $\hat{\mathcal{X}}_{4|4}$.

\newcommand{\smallwidth}{3.3in}
\newcommand{\sfigwidths}{1.5in}
\newcommand{\bigwidth}{3.1in}
\newcommand{\bfigwidths}{2.7in}
\newcommand{\legheight}{0.5in}

\begin{figure}[htb!]
    \centering
    \begin{subfigure}{\bigwidth}
         \centering
         \includegraphics[height=\legheight]{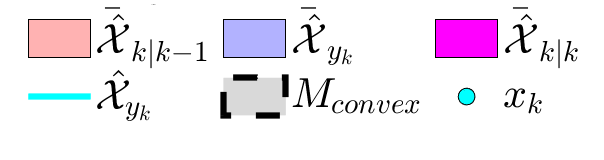}
         \label{fig:legend}
     \end{subfigure}
    \begin{subfigure}{\smallwidth}
        \begin{subfigure}[b]{\sfigwidths}
             \centering
             \includegraphics[width=\sfigwidths]{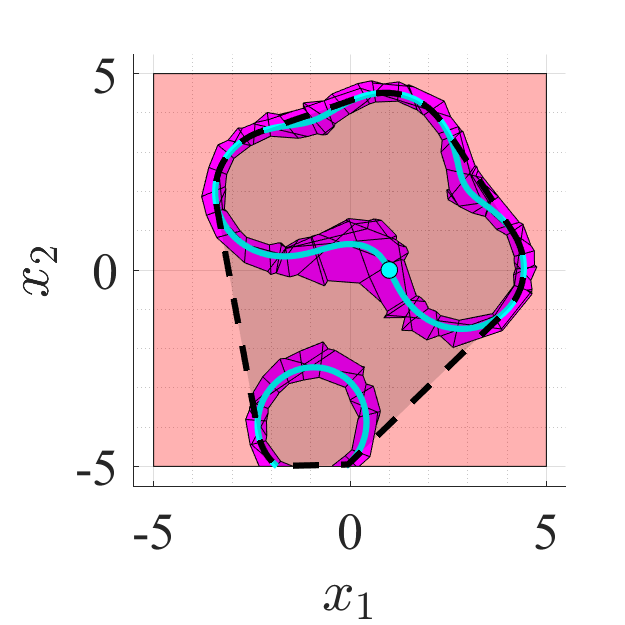}
             \caption{$k=0$}
             \label{fig:s0}
         \end{subfigure}
         \begin{subfigure}[b]{\sfigwidths}
             \centering
             \includegraphics[width=\sfigwidths]{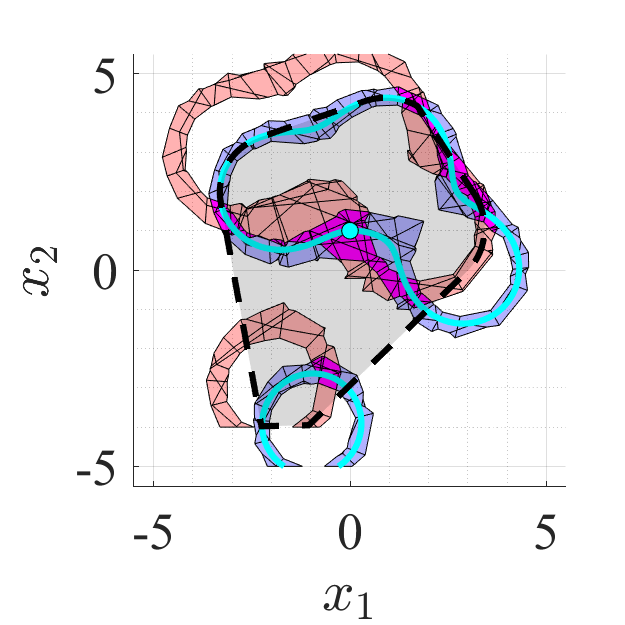}
             \caption{$k=1$}
             \label{fig:s1}
         \end{subfigure}\\
         \begin{subfigure}[b]{\sfigwidths}
             \centering
             \includegraphics[width=\sfigwidths]{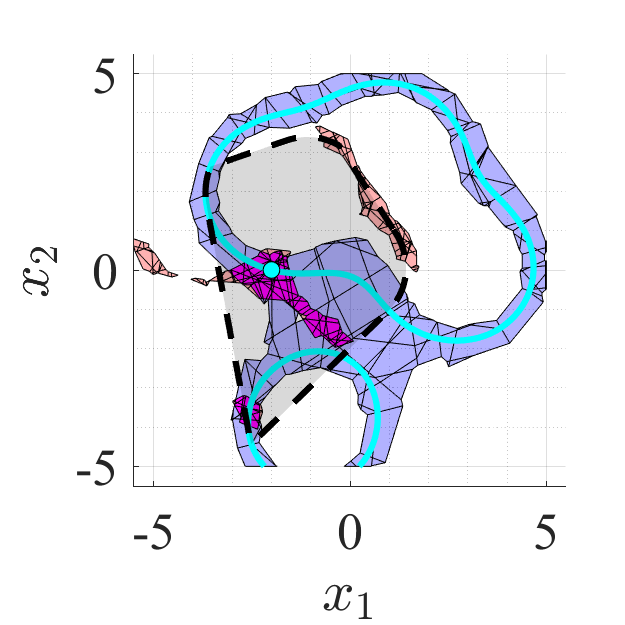}
             \caption{$k=2$}
             \label{fig:s2}
         \end{subfigure}
         \begin{subfigure}[b]{\sfigwidths}
             \centering
             \includegraphics[width=\sfigwidths]{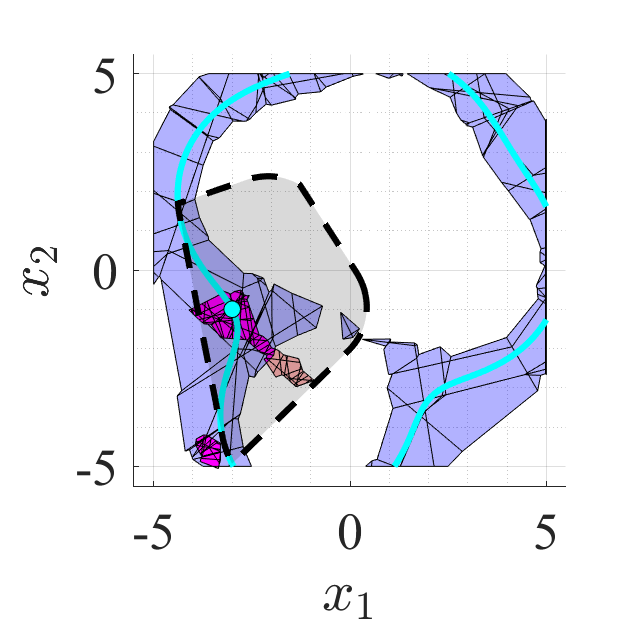}
             \caption{$k=3$}
             \label{fig:s3}
         \end{subfigure}
     \end{subfigure}
     \begin{subfigure}{\bigwidth}
         \centering
         \includegraphics[width=\bfigwidths]{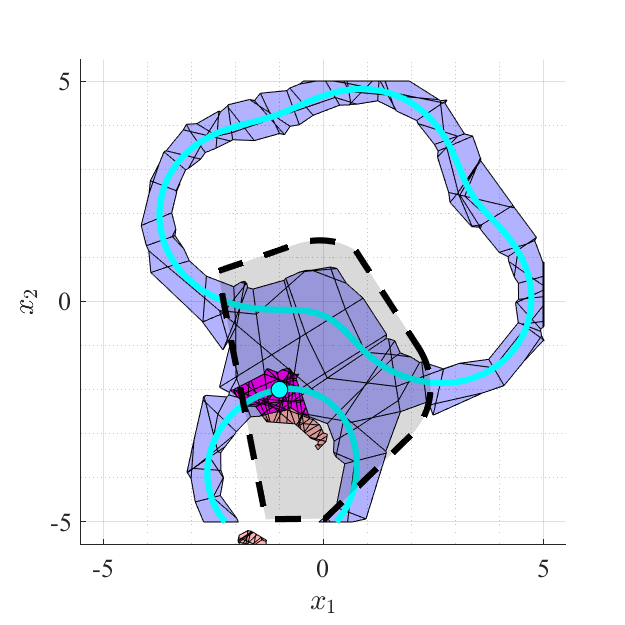}
         \caption{$k=4$}
         \label{fig:s4}
     \end{subfigure}
    \caption{Set valued state estimation \eqref{eqn-dynUpdate}-\eqref{eqn-measUpdate} of \eqref{eq:integrator} with measurement model \eqref{eqn-y-source} and its over-approximation represented as a hybrid zonotope given by Figure \ref{fig-H-tower}(b).}
    \label{fig:SVSEofSources}
\end{figure}
\addtolength{\textheight}{-1.3cm}
\section{Conclusion}
\label{sect:conclusion}

This paper presents a method for  SVSE of nonlinear, discrete-time systems with bounded input and process uncertainties and measurement noise. By combining hybrid zonotopes with graphs of functions, we efficiently obtain nonconvex set-valued state estimates. Also, several methods for generating SOS approximations of nonlinear functions are explored. A numerical example demonstrates tighter set-valued state estimates than achievable using convex methods. Future work will apply these methods to verify more computationally-efficient nonlinear observers, e.g, particle filters, similar to how reachability analysis is used to verify controllers. Techniques to mitigate growth in set memory complexity over time, e.g., periodic complexity reduction via over-approximation, should also be investigated.

\bibliographystyle{IEEEtran}
\bibliography{bibNew}

\end{document}